# Not all explicit cues help communicate: Pedestrians' perceptions, fixations, and decisions toward automated vehicles with varied appearance

Wei Lyu[a], Yaqin Cao[a], Yi Ding[a], Jingyu Li[b], Kai Tian[c,d,*], Hui Zhang[c,d]

[a] School of Economics and Management, Anhui Polytechnic University, Wuhu China, 241000

(W. L., lvw@mail.ahpu.edu.cn; Y. C., caoyaqin.2007@163.com; Y. D., yding@ahpu.edu.cn)

[b] Department of Nuclear Safety and Reliability Research, Chinergy Co., LTD，Beijing China，100193 (J. L., jingyu_li1996@163.com)

[c.] Intelligent Transportation Systems Research Center, Wuhan University of Technology, Wuhan 430063, China

[d.] Engineering Research Center of Transportation Information and Safety, Ministry of Education, China, Wuhan 430063, China (H. Z., zhanghuiits@whut.edu.cn)

*Corresponding Author: K. T., tiankai_1993@hotmail.com

## Highlights

1. Pedestrians' subjective perception, visual fixations, and road-crossing decisions towards AV were studied.

2. AVs' explicit cues were manipulated regarding driver presence, textual indication, radar system, and eHMI.

3. AVs' implicit kinematics play a dominant role in pedestrians' decisions, evidenced by multi-method metrics.

4. External textual indications and roof radars do not impact pedestrians' crossing decision time, yet require more visual resources.

5. The inclusion of an eHMI mitigates the adverse effects, in terms of reduced visual input, heightened perceived clarity, and earlier road-crossing decisions.

# Abstract

Future automated vehicles (AVs) are anticipated to feature innovative exteriors, such as textual identity indications, external radars, and external human-machine interfaces (eHMIs), as evidenced by current and forthcoming on-road testing prototypes. However, given pedestrians' vulnerability in road traffic, it remains unclear how these novel AV appearances will impact pedestrians crossing behaviour, especially from the perspective of pedestrians' multimodal performance, including subjective perceptions, gaze patterns, and road-crossing decisions. To address this gap, this study pioneers an investigation into the influence of AVs' exterior design, correlated with their kinematics, on pedestrians' road-crossing perception and decision-making. A video-based eye-tracking experimental study was conducted with 61 participants who responded to video stimuli depicting a manipulated vehicle approaching a predefined road-crossing location on an unsignalized, two-way road. The vehicle's kinematic pattern was manipulated into yielding and non-yielding, and its external appearances were varied across five types: with a human driver (as a conventional vehicle), with no driver (as an AV), with text-based identity indications, with roof radar sensors, with dynamic eHMIs adjusted to vehicle kinematics. Participants' perceived clarity, crossing initiation distance (CID), crossing decision time (CDT), and gaze behaviour, during interactions were recorded and reported. The results indicated that AVs' kinematic profiles play a dominant role in pedestrians' road-crossing decisions, supported by their subjective evaluations, CID, CDT, and gaze patterns during interactions. Moreover, the use of clear eHMI, such as dynamic pedestrian icons, reduced pedestrians' visual load, enhanced their perceptual clarity, expedited road-crossing decisions, and thereby improved overall crossing efficiency. However, it was found that both textual identity indications and roof radar sensors have no significant effect on pedestrians' decisions but negatively impact pedestrians' visual attention, as evidenced by heightened fixation counts and prolonged fixation durations, particularly under yielding conditions. Excessive visual and cognitive resource occupation suggests that not all explicit cues facilitate human-vehicle communication. The practical and safety implications of these findings for future external interaction design of AVs are discussed from the perspective of vulnerable road users.

# 1. Introduction

Pedestrians are among the most vulnerable road users (VRUs), with the World Health Organization reporting that 23% of traffic fatalities are pedestrians (WHO, 2023). In China, this number surges to 42% (Wang et al., 2019), underscoring the severity of pedestrian safety issues. One of the primary factors contributing to pedestrian fatalities is collisions with motor vehicles, with over 90% of such accidents being driver-error-related (Curry et al., 2011; Singh, 2015). To mitigate human driving errors, an important solution is to improve the level of vehicle automation. Therefore, the development of autonomous driving systems is the focus of both academia and industry and is highly anticipated by the public (Kaye et al., 2022; Lee et al., 2022; Wang et al., 2020).

Aside from the promising prospects, the introduction of AVs into road traffic has raised concerns among practitioners and researchers regarding how AVs should interact with VRUs, such as pedestrians (Dommes et al., 2021; Ezzati Amini et al., 2021; Rasouli et al., 2018; Tabone et al., 2021). Considering the disparities between AVs and conventional vehicles (CVs) in kinematics and appearances, pedestrians' road behaviour patterns may change during interactions. Specifically, AVs' less human-like driving behaviour may confuse and annoy pedestrians. Additionally, the lack of a driver role in AVs causes pedestrians to lose their communication channels with drivers (Crosato et al., 2024). Accordingly, there are arguments that breakdowns or reconstructions in communication between AVs and VRUs could affect pedestrians' trust and acceptance of AVs (Deb et al., 2018) and lead to additional traffic dilemmas and safety concerns (El Hamdani et al., 2020). The academic community believes it necessary to establish additional reliable communication channels to address these problems, which promotes much research on explicit and implicit communication between AVs and pedestrians.

## 1.1 Explicit and implicit communications

Markkula et al. (2020) defined interaction between road users as "*A situation where the behaviour of at least two road users being influenced by a space-sharing conflict between the road users*". According to this definition, a typical pedestrian-vehicle interaction scenario involves coordination, collaboration, competition, and communication between pedestrians and vehicles/drivers when pedestrians cross in front of vehicles, particularly in conflicted areas lacking traffic signals or crosswalks (Markkula et al., 2020; Predhumeau et al., 2021). Hence, pedestrians need sufficient cues from AVs to complete these complex interactive decisions. These cues are typically divided into implicit and explicit (Ackermann et al., 2019). Implicit communication cues refer to the behaviour of road users that impacts their own movement but can be interpreted as indications of intention or impending movement by other road users, such as the kinematics of vehicles. On the other hand, explicit communication cues are road user behaviours that convey signal information to others without necessarily affecting their own behaviour (Markkula et al.,

2020). Numerous studies have investigated the impact of implicit and explicit cues from AVs on pedestrians' behaviour (Ackermann et al., 2019; Dey et al., 2019; Dey et al., 2020; Madigan et al., 2023).

Regarding Implicit communication cues, numerous studies have shown the impact of vehicle distance, time to collision (TTC), and speed on pedestrian decision-making (Rasouli et al., 2018). For example, reliable evidence shows pedestrians tend to rely more on spatial than temporal distance (Song et al., 2023; Tian et al., 2022). Several studies reveal how pedestrians convert kinematic cues into visual psychophysical cues for road crossing decisions (Markkula et al., 2023; Theisen et al., 2024). Braking manoeuvres are also critical implicit information (Tian et al., 2023). Pedestrians feel comfortable and start crossing quickly when approaching vehicles slow down early and brake lightly. Harsh braking leads to pedestrian avoidance behaviour. Additionally, recent studies attempt to design active pitch movement for AVs to enhance pedestrians' perception of vehicle-yielding behaviour (Dietrich et al., 2020). Besides the longitudinal kinematical cues, Sripada et al. (2021) found that the lateral movement of AVs could help pedestrians understand AVs' intentions.

Although there appears to be a consensus that implicit cues from AVs play a dominant and primary role in interactions, particularly in experimental studies involving both types of cues (Dey et al., 2019; Dey et al., 2020; Madigan et al., 2023), this does not imply that explicit cues are merely gimmicks and not necessities (de Winter & Dodou, 2022; Tabone et al., 2021). There is growing evidence that explicit communication plays a key role in pedestrian-AV interaction, as discussed in the following section.

**1.2 Multimode novel explicit communications**

In terms of explicit communications from AVs, significant attention has been devoted to external human-machine interfaces (eHMIs) regarding their conveyed meanings, communication channels, preferred forms, colours, locations, etc. This is evident from reviews by Bazilinskyy et al. (2019), which covered 22 eHMI concepts from industry, and Dey et al. (2020), which included over 70 eHMI concepts from academia. Currently, there is no consensus on the optimal combination of various dimensions for eHMIs. A general trend suggests that clear eHMI designs (e.g., in the form of pedestrian icons displayed in AV's grille) can enhance pedestrians' understanding of AV intentions, increase their trust in AVs, and reduce their decision-making time (Guo et al., 2022). de Winter & Dodou (2022) emphasised the importance of eHMIs in future AV-pedestrian interactions, noting their potential to improve interaction efficiency, visualise vehicle intentions, and increase pedestrian acceptance. However, a recent study by Lyu et al. (2024)

highlighted that the lack of standardisation could lead to inconsistent eHMIs among AVs, thereby diminishing pedestrians' perceived clarity, prolonging their road-crossing decision times, and demanding more visual attention resources when interacting with multiple AVs on the road. This challenges the feasibility of AV's novel external appearances in complex situations.

Moreover, sound cues may enrich explicit communication information by utilising pedestrians' auditory channels. Ahn et al. (2021) found that a combination of visual and auditory cues is most effective for pedestrians' perceptions. In most cases, auditory signals facilitate pedestrian cognitive responses. A similar study from Bindschädel et al. (2023) showed that acoustic signals could achieve equal positive effects on pedestrians compared to eHMIs.

Additionally, AVs may differ in appearance from CVs. Fig. 1a-e illustrates several AV prototypes from Chinese manufacturers that have obtained open-road test licences for Level 3 AVs (SAE International., 2021). It is evident from Fig. 1a that future AVs may share similar exterior designs as CVs, such as Tesla's Model Y. AVs may also feature prominent markings to indicate their AV identity, like the text and logos (see Fig. 1b/c/d/e), especially during public testing phases. Additionally, AVs may be equipped with external sensors visible to other road users, such as roof cameras and radar (see Fig. 1c/d/e or Google's Waymo). Some AVs may also incorporate eHMIs (see Fig. 1f, or Mercedes-Benz F015, or Nissan IDS). Notably, the eHMI concept depicted in Fig. 1f has been implemented in Huawei's newly mass-produced car (Huawei IM L7).

The above research shows that in the future, pedestrians may need to meet AVs with different appearances every day. Therefore, it is necessary to understand how these differences in appearance affect pedestrians' subjective perceptions, visual patterns, and decisions.

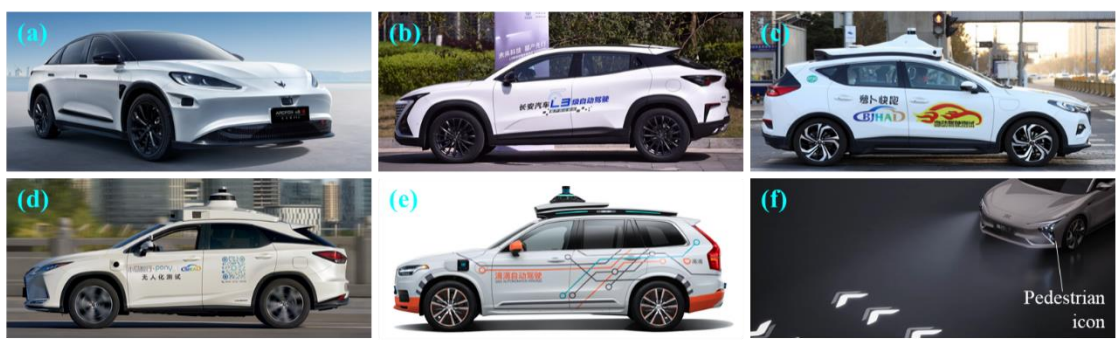

Fig. 1. Examples of AV prototypes from Chinese manufacturers with open-road test licences (a: Arcfox, https://www.arcfox.com.cn; b: Changan, http://www.globalchangan.com; c: Apollo, https://www.apollo.auto/; d: Pony, https://www.pony.ai/; e: Didi, https://www.didiglobal.com/; f: Huawei, https://www.immotors.com/)

## 1.3 Disagreements among explicit communication studies

Although the community has generally reached a consensus that eHMI may enhance pedestrian decision-making efficiency by compensating for the partially lost communication channel between AVs and pedestrians, there are certain differences of opinion in current research regarding different forms of explicit communication (Song et al., 2023).

Colley et al. (2022) found that conspicuous sensors and the distinctive appearance of AVs could enhance pedestrians' trust and understanding of AVs, while Ackermans et al. (2020) demonstrated that a distinctive appearance of AVs had no significant effect on pedestrians' overall behavioural data. Ackermans et al. (2020) further found a positive impact of distinctive appearance on participants with a more negative attitude towards AV. However, Dey et al. (2019) indicated a distinctive appearance inspired less confidence in pedestrians' crossing decisions. Additionally, although significant evidence suggested that static textual indication has no impact on pedestrian behaviour (Wilbrink et al., 2021), most AV prototypes have their autonomous driving features prominently marked, as shown in Fig. 1.

Moreover, when comparing static text to flashing eHMI, Eisma et al. (2023) found no significant difference in response time, while the text-based eHMI required attention for a longer period. However, Lau et al. (2024) showed that pedestrians initiated their crossing earlier with dynamic eHMI than no eHMI or static eHMI. Furthermore, a recent study applied recorded videos to investigate pedestrian crossing willingness when interacting with AVs equipped with light, sound, or light+sound eHMIs. They found that multimodal eHMIs were not significantly better than single-modal eHMIs. While, Ahn et al. (2021) show that a combination of visual and auditory interfaces is most effective in understanding information.

Based on the above discussion, it can be seen that there are a lot of disagreements in the current research on explicit communications. Some research results are completely inconsistent with each other, and some studies have partial disagreements. This highlights some issues in the current research on explicit communication.

### 1.4 Research gaps and questions

Most existing studies on explicit communication effects have different and relatively single independent variables. Some studies draw conclusions only through a single metric (Colley et al., 2022), while others include two types of independent variables (Bazilinskyy et al., 2020). However, crossing in front of a vehicle is a complex task coupling perception, cognition, execution, and modification. Multiple dimensions of indicators may be needed to evaluate explicit communication's impact accurately. Studies by Guo et al. (2022) and Lyu et al. (2024) on eHMIs, as part of AV's external appearance, indicated that their forms, placements, and consistency across AVs reshaped pedestrians' visual behaviour. It remains unclear whether these different exterior designs of AVs (as shown in Fig. 1), soon to be tested on public roads, would demand additional visual resources from pedestrians. In general, current studies lack a comprehensive analysis of pedestrian performance, which leads to differences in research conclusions. Moreover, existing research lacks a systematic integration of the effects of various exterior appearances of AVs (e.g., see Fig. 1) on pedestrians.

Motivated by the above research gaps, this study pioneered a systematic investigation into how the exterior design of AVs affects pedestrians' subjective perception, visual patterns, and crossing intentions. The study comprehensively considered multimodal performance metrics from pedestrians, including their gaze behaviour, road-crossing decisions, and perceived clarity, to understand the potential differences in varied exterior appearances and communication channels between AVs and CVs, including driver presence, textual indications, conspicuous sensors, and eHMIs. A video-based simulation study was conducted in which participants experienced a series of interactions with a vehicle (with its yielding patterns and external appearances manipulated) approaching a predefined road-crossing location on an unsignalised, crosswalk-free, two-way road.

## 2. Methodology

### 2.1 Participants

Approval for this experimental study was obtained from the Ethics Committee of Anhui Polytechnic University. According to G*power 3.1, with an effect size of 0.25, an α error probability of 0.05, and a power > 0.95, the minimum sample size required for this within-subject experiment was determined to be 20 participants. Recruitment advertisements were disseminated through campus social media platforms. A total of 61 participants were recruited for the experiment, consisting of 32 males and 29 females. The age range of the participants was 18 to 27 years (M = 21.590, SD = 2.772). All participants self-reported being in good health, with no head conditions affecting their cognitive decision-making or limb conditions affecting daily mobility. Additionally, all participants were right-handed and had normal or corrected-to-normal vision. Informed consent was obtained from all participants before the experiment, and they were compensated with 50 RMB upon completion.

### 2.2 Apparatus and materials

#### 2.2.1 Apparatus

This pedestrian-vehicle interaction experiment, based on a video-based simulation paradigm (Lyu et al., 2024; Zhao et al., 2023), was conducted at the Human Factors and Ergonomics

Laboratory of Anhui Polytechnic University. The experimental procedure was designed using E-Prime 3.0 software (Psychology Software Tools, USA) and presented using a Tobii Pro Spectrum eye tracker (Tobii AB, Sweden), as shown in Fig. 2. E-Prime 3.0 was utilized to design the experimental program, control the experimental process, and send event markers to the eye tracker to ensure synchronized timestamps. Additionally, E-Prime 3.0 recorded participants' behavioural responses (e.g., reaction time) and subjective evaluations (e.g., Likert scale ratings) during the stimuli presentation. The Tobii Pro Spectrum displayed a series of video stimuli on an accompanying screen (size: 23.8 inches, resolution: 1920 × 1080 pixels) and recorded participants' eye movements while screening video stimuli at a sampling rate of 600Hz.

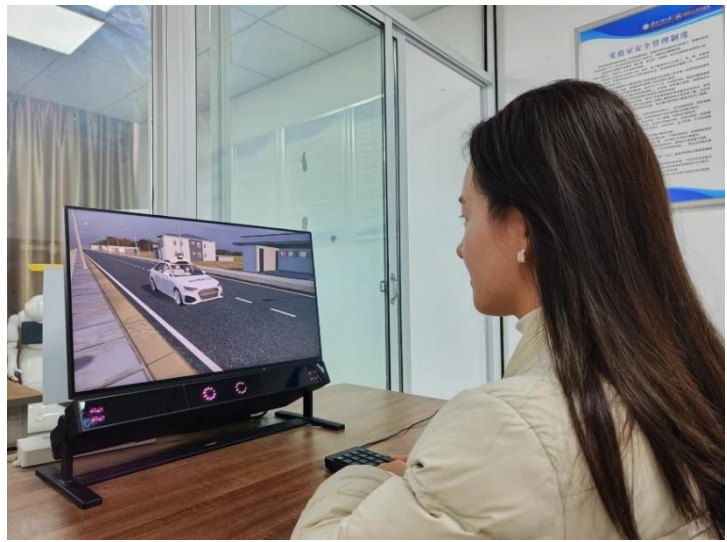

Fig. 2. Experiment scenario on the eye-tracker platform

**2.2.2 Materials**

The experiment materials consisted of 50 videos, which were obtained by repeating 10 original videos 5 times, generated with the Unity 3D software (Unity Technologies, USA). Each video was approximately 19 seconds long, with a resolution of 1980 × 1080 pixels and a playback frame rate of 30 fps. The videos depicted the traffic scenario where a pedestrian is waiting to cross the street while a car (either controlled by a human driver or an automated system) approaches from their left. The experiment utilized a suburban traffic scene with two-way, two-lane roads, devoid of traffic lights and crosswalks. The road width was set at 3.5 meters, and the car's speed limit was 50 km/h, in accordance with Chinese road standards. Additionally, these videos were presented from the pedestrian's perspective, with horizontal and vertical viewing angles of 150 degrees and 130 degrees, respectively, and an eye height of 160 cm, based on the National Standard of China: Human Dimensions of Chinese Adults (GB/T 10000-2023).

## 2.3 Experimental design

### 2.3.1 Independent variables

This experiment employed a two-factor, within-subject design. Two independent variables, namely the yielding pattern and external appearance of the approaching vehicle, were manipulated.

**(1) Yielding patterns**

Overall, the kinematic behaviour of the approaching vehicle was categorised into non-yielding and yielding, based on previous research (Dey et al., 2019; Lyu et al., 2024; Tian et al., 2023).

In the non-yielding condition (as shown in Fig. 3a), the vehicle appeared 80 m away and maintained a constant speed of 50 km/h throughout its approach. At the 3-second mark (with a travelling distance of 41 m), the vehicle may display a dynamic eHMI in the form of a flashing AV icon to convey its non-yielding intention, if applicable. The vehicle then proceeded to pass the predefined pedestrian waiting location at a constant speed. Following this, the traffic scene remained for about 13 s to allow for the possibility of the pedestrian initiating a crossing after the vehicle passed.

In the yielding condition (as shown in Fig. 3b), the car appeared 80 meters away and travelled 41 meters at a speed of 50 km/h (with an elapsed time of 3 s). Subsequently, the car began to decelerate with an acceleration of -2.78 m/s. Simultaneously, the car may display a dynamic eHMI in the form of flashing pedestrian icons to convey its yielding intention, if applicable. After covering a braking distance of 35 m (with an elapsed time of 5 s), the vehicle stopped 4 m away from the pedestrian. These kinematic parameters are similar to previous studies (Lyu et al., 2024; Zhao et al., 2023). Following this, the traffic scene remained for approximately 11 s to account for the possibility of the pedestrian initiating a crossing after the vehicle stopped.

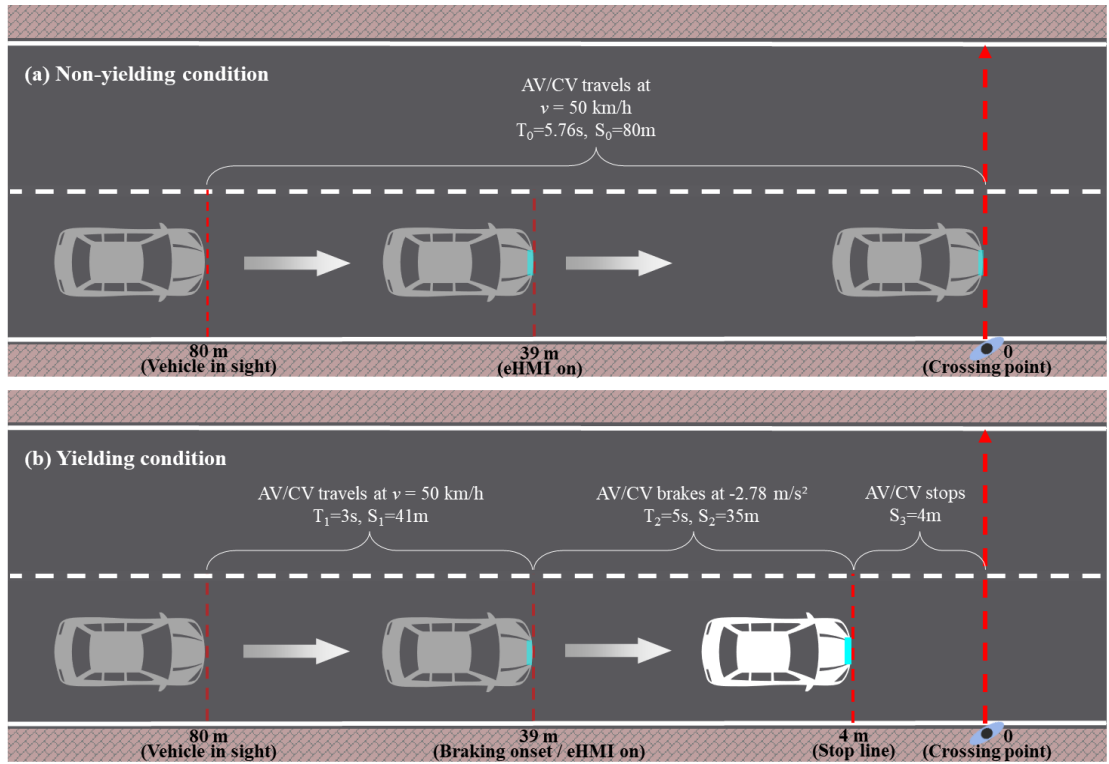

Fig. 3. Scheme of experimental design under non-yielding and yielding condition

（2）**External appearances**

Five different external appearances were devised for both yielding and non-yielding vehicles, motivated by current road-testing initiatives (see Fig. 1) and AV prototypes from academia. These external appearance manipulations are delineated in Fig. 4.

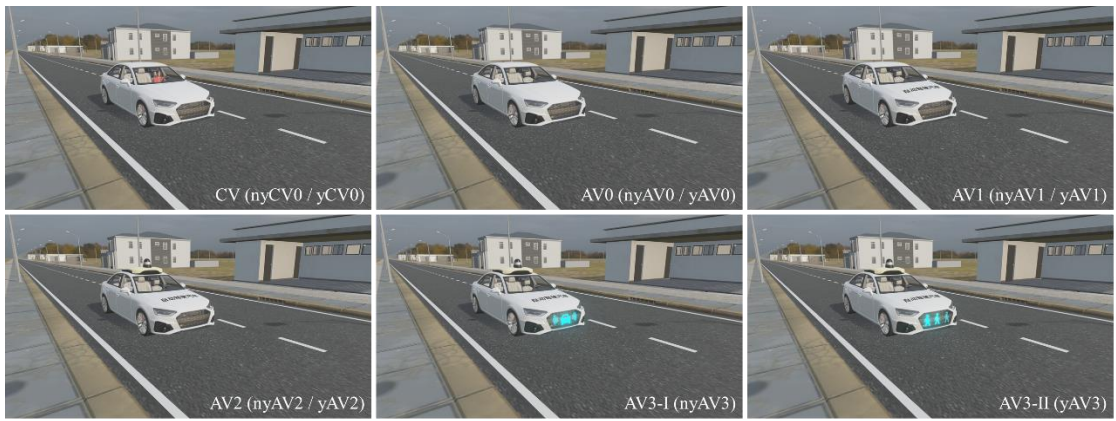

Fig. 4. Screenshots of appearance manipulation for non-yielding (ny) and yielding (y) vehicles

**CV0**: A conventional vehicle with a visible human driver, devoid of any additional external manipulations, serving as the comparative baseline.

**AV0**: An automated vehicle with no human driver and lacking additional external manipulations, akin to Fig. 1a or Tesla's Autopilot.

**AV1**: Building upon AV0, this variant features textual indications of its automated driving identity, referring to Fig. 1b or Rodríguez Palmeiro et al. (2018).

**AV2**: Extending from AV1, this prototype incorporates a conspicuous radar sensor mounted on AV's roof, as depicted in Fig. 1c/d/e or Ackermans et al. (2020).

**AV3**: Further augmenting AV2, an eHMI is situated in the grille position of the vehicle. For non-yielding AVs, an eHMI in the form of a flashing AV icon was chosen to convey the intention of not yielding, as referenced in Joisten et al. (2020). Conversely, for yielding AVs, a dynamic pedestrian-symbolic eHMI was selected to convey the intention of yielding, following the findings of Lyu et al. (2024). The colour of the eHMI was set to cyan (RGB = (0, 255, 255)), with a pulsing frequency of 0.5Hz, as shown in Guo et al. (2022).

The manipulations described above, pertaining to both kinematic profile and external appearance, collectively generated 10 distinct video stimuli. Each video was repeated five times during the experimental procedure, resulting in a total of 50 trails.

**2.3.2 Dependent variables**

The following dependent variables and metrics in this pedestrian-vehicle interaction experiment were recorded and reported:

**Crossing initiation distance (CID).** This refers to the distance between the approaching vehicle and the pedestrian at the moment they commenced road-crossing, marked by pressing the "Enter" key on a keyboard as an indicator of road-crossing initiation. Herein, the CID metric is similar to the $Z_c$ proposed by Tian et al. (2023) in their pedestrian-AV interaction research.

**Crossing Decision Time (CDT)**. This denotes the duration elapsed from the appearance of the vehicle to pedestrians' road-crossing initiation, which represents pedestrians' crossing decision time.

**Clarity Rating Score (CRS)**. Participants provided subjective ratings regarding the clarity of the message conveyed by the vehicle when they intended to cross the road. Ratings were collected after each trial on a Likert scale ranging from 1 (completely disagree) to 9 (completely agree) (Lyu et al., 2024).

**Gaze-based Metrics**. For each trial, dynamic Areas of Interest (AOIs) were delineated 3 s after the video presentation (i.e., when different manipulations began to display) using Tobii Pro Lab, a supporting software for the Tobii eye tracker. Specifically, 4 AOIs covering the grill, hood, windshield, and roof of the vehicle were designated (see Fig. 5). This approach aligns with our external appearance manipulation (see Fig. 4). Within the predefined AOIs, two commonly used metrics were extracted: Number of fixations and Average duration of fixations (Dey et al., 2019; Guo et al., 2022; Lyu et al., 2024). The number of fixations is considered highly correlated with the importance and level of interest of the AOI, while the average duration of fixations is typically used to reflect individuals' engagement and effort within the AOIs, positively correlating with processing workload (Mahanama et al., 2022; Tobii., 2021).

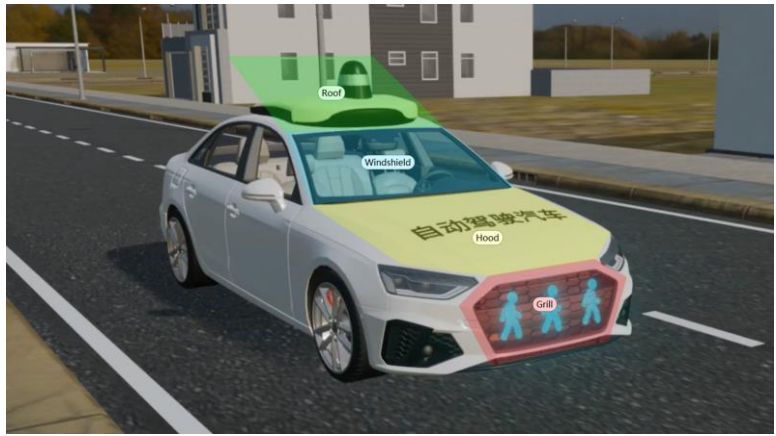

Fig. 5. The distribution of Areas of Interest in the experiment

**Subjective experience from a post-experiment questionnaire.** After the experiment, participants were surveyed regarding their general trust in and attitudes towards AVs, and interaction experience during the experiment. The questionnaire comprised three parts, totalling 24 items. The first part collected demographic information, including 7 items covering gender, age, education level, and driving experience. The second part assessed participants' general trust in AVs using 5 items adapted from Palmeiro et al. (2018) and their attitudes toward crossing in front of AVs using 4 items adapted from Deb et al. (2017). Participants responded using a five-point Likert scale, where 1 represented "strongly disagree" and 5 represented "strongly agree". Details are presented in Table 1. The third part investigated factors influencing participants' road-crossing decisions when interacting with AVs.

Table 1 Post-experiment questionnaire investigating pedestrians' general trust and attitude towards AVs

| Dimension | Coding | Items | Response |
|---|---|---|---|
| General trust | GT1 | In general, I trust AVs. | [1] [2] [3] [4] [5] |
| | GT2 | I trust AVs to avoid obstacles. | [1] [2] [3] [4] [5] |
| | GT3 | I trust AVs to keep the right lane. | [1] [2] [3] [4] [5] |
| | GT4 | I trust AVs to interact safely with pedestrians. | [1] [2] [3] [4] [5] |
| | GT5 | AVs will enhance the overall transportation system. | [1] [2] [3] [4] [5] |
| Attitude | AT1 | I would feel safe to cross roads in front of AVs. | [1] [2] [3] [4] [5] |
| | AT2 | I would feel pleasant to cross roads in front of AVs. | [1] [2] [3] [4] [5] |
| | AT3 | I would observe surroundings less to cross roads in front of AVs. | [1] [2] [3] [4] [5] |

| | AT4 | It would take less mental effort to cross roads in front of AVs. | [1] [2] [3] [4] [5] |

## 2.4 Participants' task and procedure

### 2.4.1 Participants' task

Upon arrival, participants were asked to sign the informed consent form, which provided detailed descriptions of the experimental scenarios and tasks, as follows:

"Dear participant! Welcome to the pedestrian-vehicle interaction and road-crossing experiment. During the experiment, you will view a series of videos depicting approaching vehicles. You will be standing by the side of a two-way road, waiting to cross. A vehicle will approach from your left side. This vehicle may be controlled by a human driver or by an automated driving system (without a driver). Additionally, the vehicle may have novel external designs to convey relevant information. Due to the absence of traffic lights or crosswalks, the approaching vehicle may or may not yield to you. Based on the scenario information and your experience in real traffic, when you believe it is safe to cross the road, press the 'ENTER' key to indicate your decision to initiate crossing. Following this, please respond to a single question about your perceived clarity using the numeric keys 1-9. Subsequently, there will be a brief blank screen period, after which the next trial will commence."

### 2.4.2 Procedure

Afterwards, participants were guided to sit in front of the eye tracker, maintaining a distance of 60 cm from the monitor, and keeping their gaze centred on the screen. The experimenters then initiated the experimental procedure using E-Prime 3.0 (as illustrated in Fig. 6).

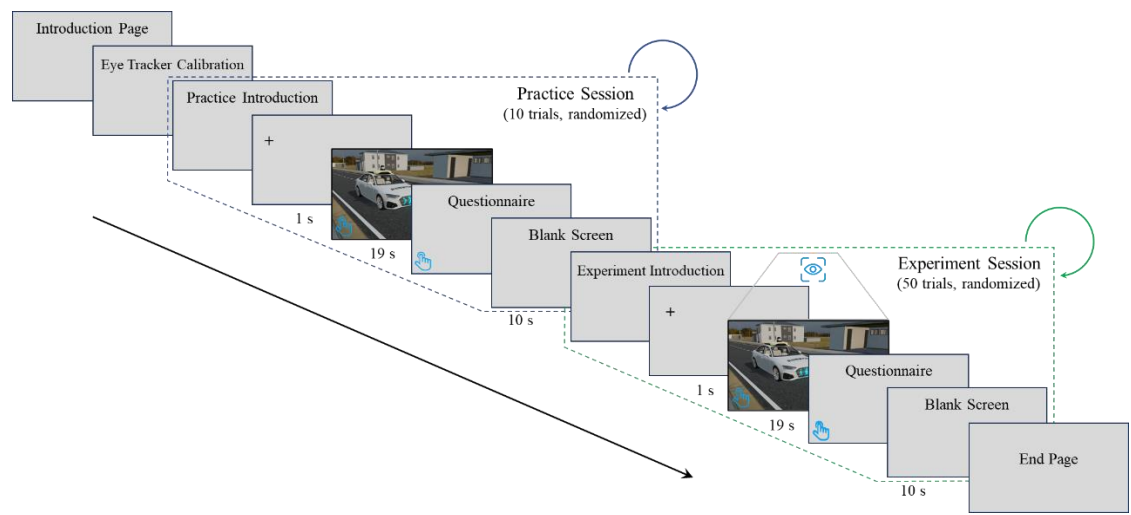

Fig. 6. The experiment procedure

**(1) Introduction page**. This page presented the introduction to the experimental scenarios and tasks. Participants were allowed to ask questions.

**(2) Eye tracker calibration**. A 9-point calibration protocol was used to ensure the accuracy and precision of the eye tracker. Participants were instructed to remain still and follow the movement of a small animated object on the screen.

**(3) Practice session**. This session included 10 practice trials to familiarize participants with the experimental procedure and tasks.

**a. Practice introduction**. Participants were informed that the practice session was about to begin. They could ask questions to the experimenters if needed during this session.

**b. Blank screen with a '+'**. This drift correction step used a '+' to direct participants' attention to the location where the vehicle would soon appear.

**c. Video stimulus presentation**. Participants watched video stimuli and made road-crossing decisions by pressing 'ENTER' when they deemed it safe. The practice session included 10 randomly presented trials (5 non-yielding and 5 yielding trials). For yielding AVs with an eHMI, a flashing arrow eHMI was used, as referenced in Guo et al. (2022).

**d. Perceived clarity questionnaire**. Participants rated their perceived clarity of the information conveyed by the vehicle in the current trial using numeric keys 1-9.

**e. Blank screen**. A brief interlude before the next trial.

**(4) Experimental session**. This session comprised 50 trials where participants followed the procedures and tasks outlined in the practice session. During this session, E-Prime 3.0 sent beginning and ending markers for each trial to the Tobii eye tracker.

**(5) End page**. Participants were notified of the end of the experiment.

After the experiment, participants received monetary compensation and were reminded to complete the online questionnaire as soon as possible.

## 2.5 Data preparation and analysis

During the experiment, one participant failed to complete the experiment and data collection due to a software malfunction. As a result, data from 3000 trials were collected from a total of 60 participants.

### 2.5.1 Road-crossing data preparation

According to the experimental design (see Fig. 3), during the first 3 s of each video stimulus,

all AVs exhibited identical driving behaviour and external appearances. In other words, the approaching vehicles began to display different yielding patterns and eHMI (if applicable) after 3 s. Participants were instructed to initiate crossing when they felt it was safe, which means that they had the discretion to decide when to commence crossing. During the experiment, it was observed that in some trials (118 out of 3000 trials), participants initiated crossing within 3 seconds. Such early-crossing phenomena have also been reported in other pedestrian-vehicle interaction studies (Lee et al., 2022; Lyu et al., 2024; Tian et al., 2023). In these trials, participants' crossing decision times were less than 3 s, occurring before the experimental manipulations were presented. Consequently, for the subsequent data analysis of crossing decisions, these early-crossing trials were included only in the descriptive analysis and excluded from the inferential statistical analysis.

**2.5.2 Gaze data preparation**

Regarding the eye-tracking data, despite using a rigorous calibration procedure and a static sitting posture to ensure the accuracy and precision of gaze data, several technical, environmental, or experimental factors still affected the collection of valid data. Before analysis, the eye-tracking data underwent the following checks and filtering.

**(1) Exclusion based on gaze sample rate.** Participants with a gaze sample rate below 70% throughout the experiment were excluded (Bindschädel et al., 2022)(Lyu, Zhang, et al., 2024), which resulted in the exclusion of 5 participants, corresponding to 250 trials.

**(2) Exclusion of trials without recorded fixations in predefined AOIs**. A total of 252 trials, where no fixations were recorded in the four predefined AOIs, were excluded. This could be due to participants making early crossing decisions (less than 3 s) before the dynamic AOIs appeared. It could also be due to participants focusing on other areas outside the AOIs, such as vehicles' wheels, road surfaces, or other environmental elements (Lévêque et al., 2020). These areas were not investigated and manipulated in this study and were therefore not defined as target AOIs.

**(3) Exclusion of Roof AOI from statistical analysis.** Based on the locations of experimental manipulations and previously identified pedestrian visual patterns, we defined four dynamic AOIs on the vehicle: Grill, Hood, Windshield, and Roof. However, preprocessing of the experimental data revealed that fixations in the Roof AOI were recorded in only 8.12% of the trials, compared to over 80% of the trials for the other AOIs. Due to the significant difference in sample sizes, subsequent statistical analyses did not include the Roof AOI data, which were only provided in descriptive analyses where necessary.

**2.5.3 Statistical analysis**

The pre-processed data were analysed using SPSS 27.0 (IBM SPSS Inc, USA). The normality of each metric was assessed using the Kolmogorov-Smirnov test. Following the homogeneity of variance test, analysis of variance (ANOVA) was conducted for statistical analysis. Fisher's LSD post-hoc test was employed to compare differences when the main effect was significant. An α level of 0.05 was used for all statistical tests. Effect sizes and statistical power were computed and reported.

# 3. Results

In this section, we first report the findings from the participants' post-experiment questionnaire, which includes their general trust in and attitudes towards AVs, as well as the factors influencing their road-crossing decisions during interactions with vehicles. Next, we present a descriptive analysis of participants' crossing behaviours, including their crossing initiation distance (relative to the vehicle's position) and crossing decision time. Following this, we detail the statistical analysis results for crossing decision time, perceived clarity, and eye-tracking metrics under yielding and non-yielding conditions.

## 3.1 Post-experiment questionnaire

Fig. 7 presents the survey results of participants' general trust in and attitudes toward AVs. On average, nearly 50% of participants expressed trust in AVs, particularly in their lane-keeping capabilities (GT3, 66%) and their overall enhancement of the traffic system (GT5, 64%). About 11% of participants expressed distrust in AVs, with the same proportion believing that AVs cannot interact safely with pedestrians.

When it comes to attitudes toward crossing in front of AVs, the outlook is less optimistic. Only 45% of participants believed it was safe to cross in front of an AV, while the remaining participants were either neutral (38%) or opposed (17%). Regarding whether crossing in front of an AV reduces mental effort or attention to surroundings, the proportion of participants in favour and those against was roughly equal (~40%), with slightly more participants expressing opposition.

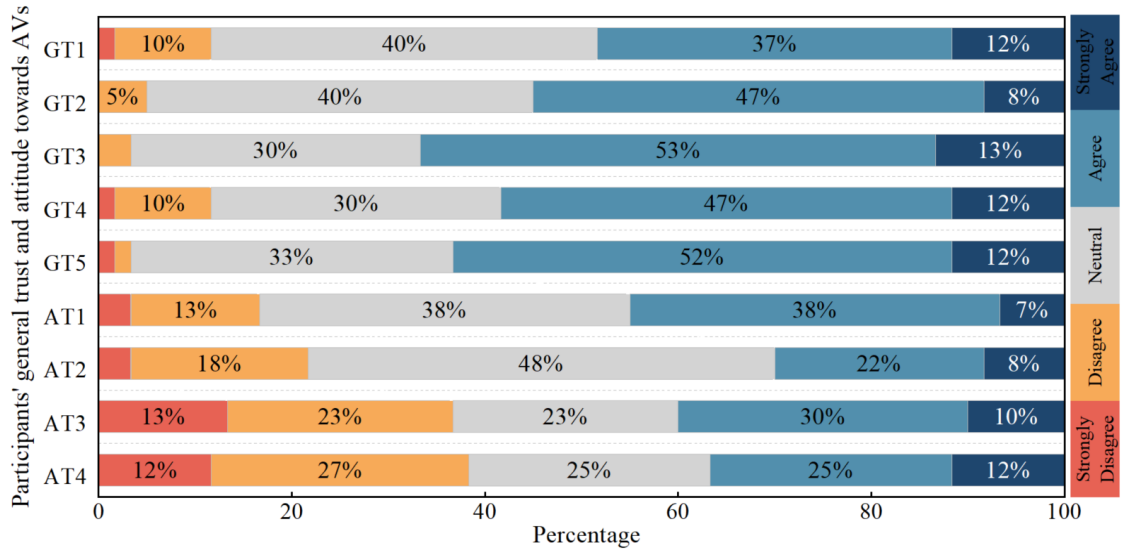

Fig. 7. Participants' general trust and attitude towards AVs

The post-experiment questionnaire also investigated factors influencing pedestrians' road-crossing decisions, as shown in Fig. 8. Over 95% of participants considered the speed of vehicles (PE1) and yielding behaviour (PE4) to be important factors in their road-crossing decisions. Additionally, other implicit cues from vehicles, such as the spatial gap (PE2, 85%) and temporal gap (PE3, 69%) between the pedestrian and the vehicle were also assigned high importance. Concerning explicit cues, over 65% of respondents deemed the textual indication of AV and eHMIs to be important for their road-crossing decisions. However, over 50% of respondents considered the suspicious radar device on the AV's roof to be unimportant or neutral in their road-crossing decisions. Furthermore, regarding the presence of a human driver, only 41% of respondents deemed it important, while 28% considered this factor to be unimportant in their road-crossing decisions.

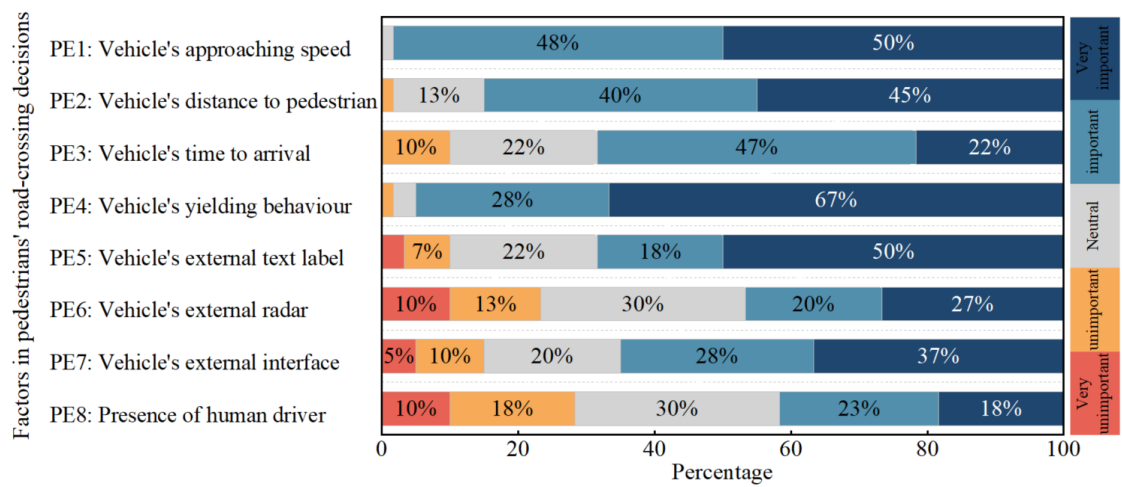

Fig. 8. Importance of factors in pedestrians' road-crossing decisions

## 3.2 Descriptive analysis of pedestrians' road-crossing behaviour

**(1) Crossing initiation distance (CID)**

Fig. 9 illustrates the distribution of participants' CID under different manipulations of the independent variables. It is evident from Fig. 9 that participants exhibited distinctly different behavioural patterns depending on whether the vehicle yielded (see the cyan bars in Fig. 9f-j) or did not yield (see the yellow bars in Fig. 9a-e).

When the approaching vehicle did not yield, the majority of participants (92.13%) chose to cross after the vehicle passed. The CID distribution in this scenario was relatively flat, with an average CID of 20.61 m from the departing vehicle. Notably, when non-yielding AVs were equipped with an eHMI in the form of a flashing AV icon, participants' average CID were shorter, and the proportion of participants crossing in front of the vehicle slightly increased (see Fig. 9e).

In contrast, when the approaching vehicle yielded, over 90% of participants initiated crossing in front of the vehicle when it had either completely stopped or was close to stopping, with an average CID of 7.31 m. It is important to note that the yielding behaviour of the vehicle was predefined as stopping 4 m away from the pedestrian. When yielding AVs displayed an eHMI in the form of sweeping pedestrian icons, participants' average CID increased (9.69 meters, compared to 6.17-7.13 m in other yielding groups), indicating that participants initiated crossing from a greater distance as the vehicle approached (see Fig. 9j).

Additionally, Fig. 9 shows that in some trials (118 out of 3000), participants triggered the crossing button when the vehicle just appeared and was far away (e.g., more than 40 m), at a point where the vehicle had not yet displayed yielding or non-yielding intentions.

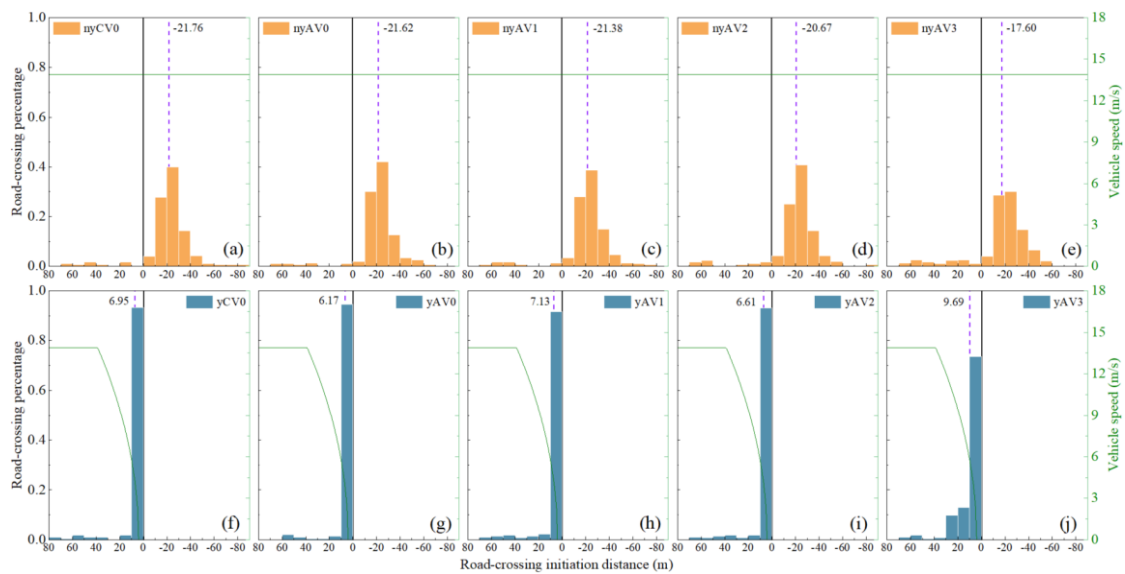

Fig. 9. The distribution of pedestrians' road-crossing percentage over road-crossing initiation

distance under different manipulations. The black solid line marks pedestrians' location, while the purple dashed line indicates the average CID. Positive values of the CID represent the vehicle approaching from the left side of the pedestrian, and negative values indicate the vehicle passing to the right side of the pedestrian.

**(2) Crossing decision time (CDT)**

Fig. 10 presents the histogram of the distribution of participants' CDT under different combinations of variable manipulations. In the non-yielding condition (see the yellow bars in Fig. 10a-e), the average CDT is M = 7.472s (SD = 1.062). A closer look at participants' CDT indicates that for 92.13% of all valid trials, their CDT exceeded 5.76 s which corresponds to the time it takes for a non-yielding vehicle to pass the pedestrian's position (see Fig. 3a). This reveals that the vast majority of participants initiated crossing from behind the vehicle within a relatively concentrated time frame after the vehicle had passed.

In the yielding condition (see the blue bars in Fig. 10f-j), the average CDT is 8.399 s (SD = 1.721), with a more evenly distributed range. The application of an eHMI appears to expedite the decision-making process, reducing the pedestrians' CDT by approximately 1 s.

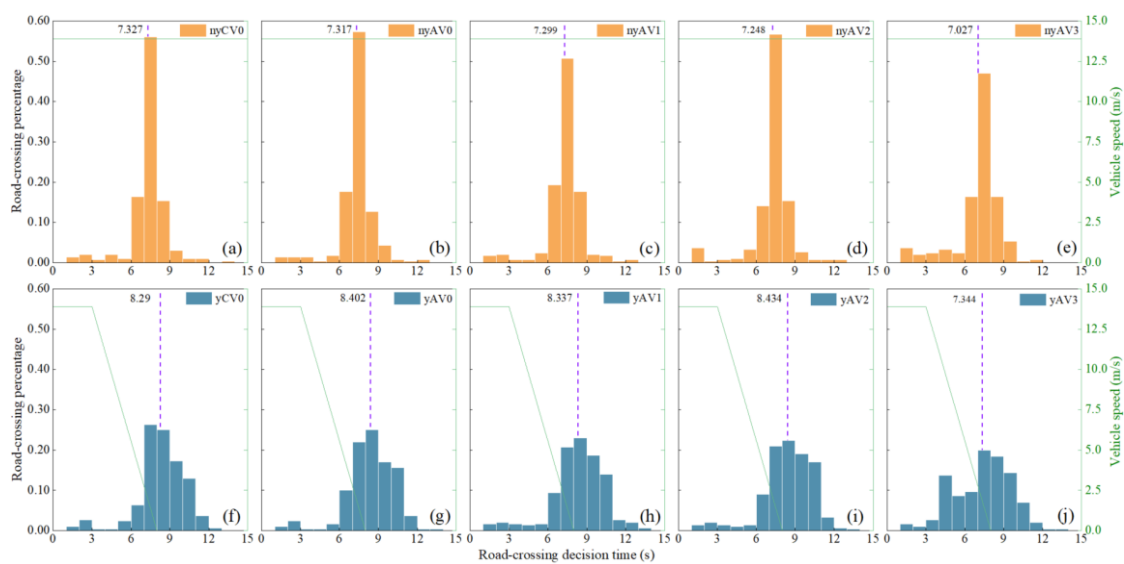

Fig. 10. The distribution of pedestrians' road-crossing percentage over road-crossing decision time under different manipulations. The purple dashed line indicates the average CDT.

Similar to Fig. 9, Fig. 10 also reveals that a portion of participants made early road-crossing decisions, with their CDT<3s (totalling 118 trials). Data from these trials were not included in subsequent statistical analyses. Furthermore, due to the notable differences in road-crossing patterns between yielding and non-yielding conditions, their perceived, behavioural, and gazed-based metrics were reported separately across these two conditions (Lee et al., 2022; Lyu et al., 2024).

## 3.3 Statistical analysis of metrics under yielding condition
**(1) Crossing decision time (CDT) and clarity rating score (CRS)**

Under the yielding condition, a one-way MANOVA was conducted to examine the effect of external appearance manipulations on CDT and CRS, as shown in Table 2. The results indicate a significant main effect of appearance manipulations on CDT ($F(4, 1439) = 25.506$, $p < 0.001$, $\eta p^2 = 0.066$). Further post-hoc comparisons with the LSD test (as shown in Fig. 11a) reveal that the AV with a pedestrian icon-based eHMI (yAV3) led to significantly shorter CDT (M = 7.518s, SE = 0.118), exceeding a difference of 1 s compared to the other four groups (see Fig. 11a).

Additionally, the main effect of appearance manipulations on CRS was also significant ($F(4, 1439) = 49.718$, $p < 0.001$, $\eta p^2 = 0.121$). Pairwise comparisons with the LSD test showed that yAV3 received the highest perceived clarity score (8.732 out of 9), which was significantly higher than the other four groups. In contrast, the vehicle with only textual labels indicating its AV identify (yAV1) received a relatively lower perceived clarity score.

Table 2 The ANOVA results for CDT and CRS under yielding condition

| Metrics | External appearance | Descriptives | | | ANOVA | | | |
|---|---|---|---|---|---|---|---|---|
| | | Sample size | Mean | S.E. | *df* | *F* | *p* | *ηp²* |
| CDT | yCV0 | 286 | 8.607 | 0.084 | (4, 1439) | 25.506 | <0.001 | 0.066 |
| | yAV0 | 289 | 8.619 | 0.085 | | | | |
| | yAV1 | 288 | 8.605 | 0.101 | | | | |
| | yAV2 | 290 | 8.652 | 0.096 | | | | |
| | yAV3 | 291 | 7.518 | 0.118 | | | | |
| CRS | yCV0 | 286 | 7.472 | 0.101 | (4, 1439) | 49.718 | <0.001 | 0.121 |
| | yAV0 | 289 | 7.222 | 0.101 | | | | |
| | yAV1 | 288 | 7.097 | 0.110 | | | | |
| | yAV2 | 290 | 7.372 | 0.098 | | | | |
| | yAV3 | 291 | 8.732 | 0.046 | | | | |

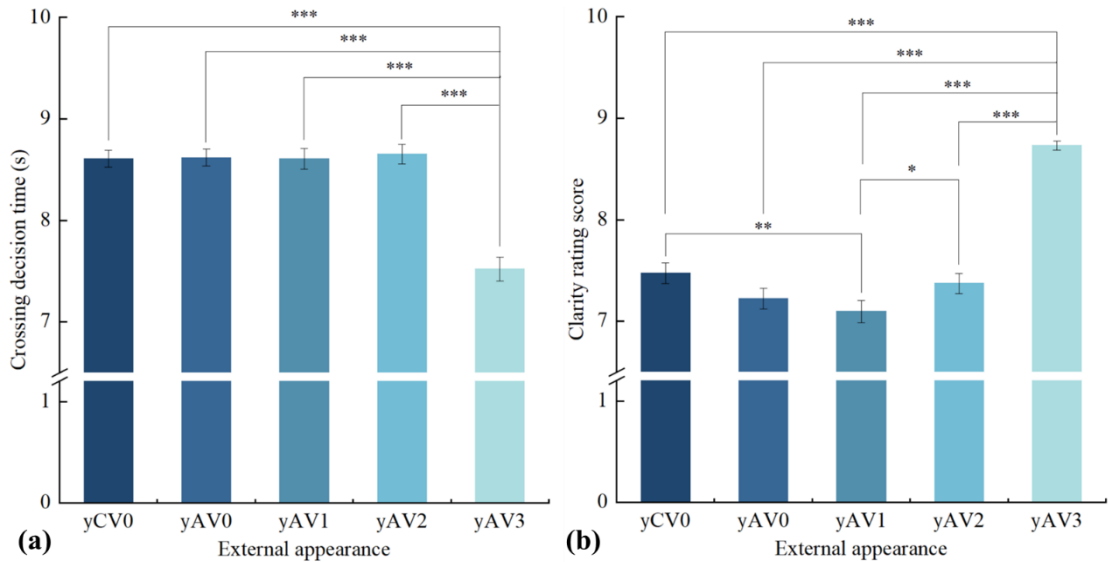

Fig. 11. Post-hoc comparisons of CDT (a) and CRS (b) under yielding condition, where the error bar depicts standard error, * depicts p < 0.05, ** depicts p < 0.01, *** depicts p < 0.001

**(2) Gaze-based metrics**

Considering the external appearance manipulations and designated AOIs involved in this study, a two-way MANOVA was conducted to examine the effects of external appearance manipulations and AOI locations on participants' Number of fixations and Average duration of fixations during interaction with yielding vehicles. In cases where main effects or interaction effects were significant, LSD post-hoc tests were used for multiple comparisons. For both metrics—Fixation counts and Average duration of fixations—significant main effects and interaction effects were found for external appearance and AOI location, as shown in Table 3.

Table 3 The ANOVA results for gaze-based metrics under yielding condition

| Gaze Metric | Source | df | F | p | $\eta p^2$ |
|---|---|---|---|---|---|
| Number of fixations | External appearance | 4 | 7.118 | < 0.001 | 0.009 |
| | AOI Location | 2 | 29.080 | < 0.001 | 0.018 |
| | External appearance * AOI location | 8 | 16.958 | < 0.001 | 0.041 |
| Average duration of fixations | External appearance | 4 | 21.745 | < 0.001 | 0.027 |
| | AOI Location | 2 | 18.133 | < 0.001 | 0.011 |
| | External appearance * AOI location | 8 | 13.198 | < 0.001 | 0.032 |

For the metric of Number of fixations, Fig. 12a and Fig. 12b present the LSD pairwise comparison results for external appearance and AOI location, respectively. Regarding external appearance, yAV1 and yAV2 received the highest number of fixations, while participants fixated the least on yAV3 and yCV0. For the AOI location, the number of fixations on the Grill and Hood areas was significantly higher than on the windshield area.

Fig. 12c shows the distribution of fixations across different AOIs when participants interacted with vehicles under different external appearance manipulations. In the Grill area, yAV3 received the most fixations (p<0.001), while yCV0 received the fewest (p<0.05). In the Hood area, yAV1 and yAV2 had significantly more fixations compared to yCV0, yAV0, and yAV3 (p<0.001). In the Windshield area, yAV3 had the fewest fixations (p<0.01).

For the metric of Average duration of fixations, Fig. 12d and Fig. 12e present the LSD pairwise comparison results for external appearance and AOI location, respectively. For external appearance, participants had the longest fixation durations on yCV0, significantly longer than all other groups (p<0.001), followed by yAV1 and yAV2, with yAV3 having the shortest fixation durations. For the AOI location, the fixation duration on the Hood and Windshield areas was significantly higher than on the Windshield AOI. Fig. 12f shows the distribution of fixation durations across different AOIs for vehicles with different external appearances. In the Grill area, there were no significant differences in fixation duration among the different external appearances. In the Hood area, yAV3 and yAV0 had shorter fixation durations compared to the other groups. In the windshield area, yCV0 had significantly longer fixation durations than all AVs, and yAV3 had significantly shorter fixation durations compared to other vehicles without eHMI (p<0.01).

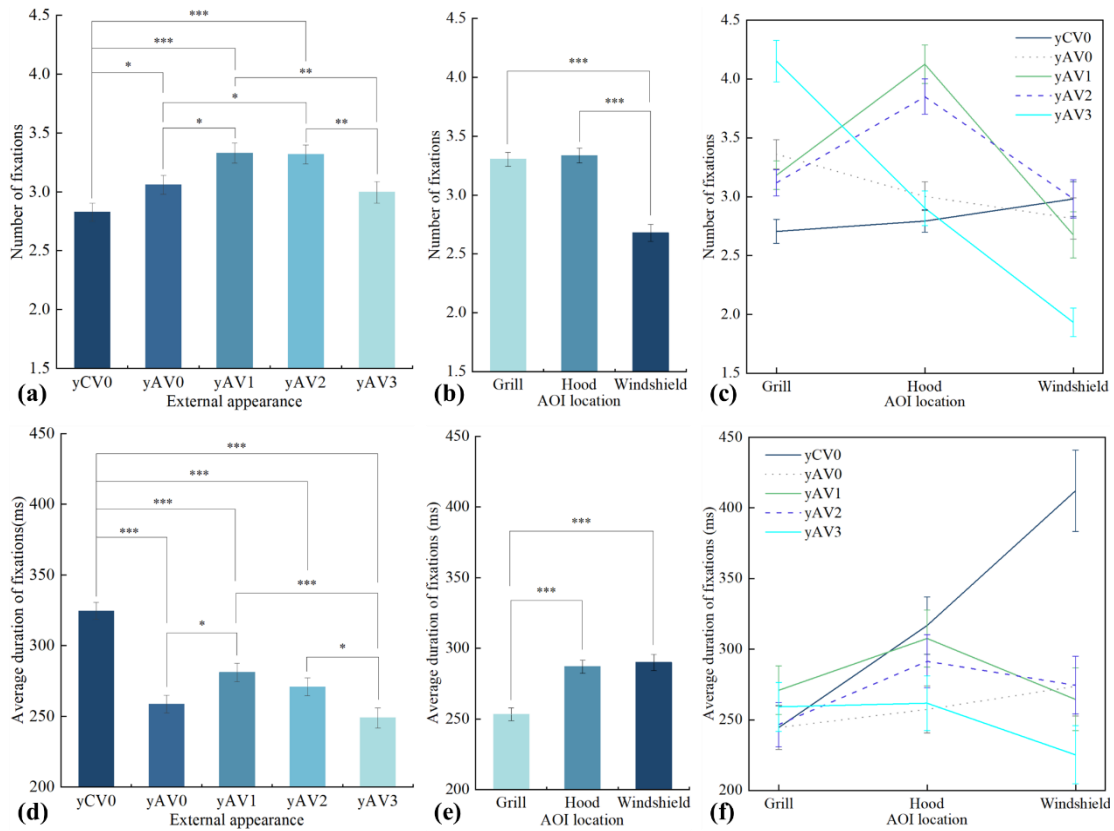

Fig. 12. Post-hoc comparisons and interaction effect for gaze-based metrics under yielding condition

## 3.4 Statistical analysis of metrics under non-yielding condition

**(1) Crossing decision time and clarity rating score**

Under the non-yielding condition, a one-way MANOVA was conducted to examine the effect of vehicle's external appearance on CDT and CRS. The results (as shown in Table 4) indicated that the main effect on CDT was not significant ($F(4, 1433) = 1.445$, $p = 0.217$, $\eta p^2 = 0.004$); however, the main effect on CRS was significant ($F(4, 1433) = 24.882$, $p < 0.001$, $\eta p^2 = 0.065$). The LSD post hoc comparisons revealed that participants' perceived clarity under the nyAV3 condition (M = 7.351, S.E. = 0.152) was significantly higher than in the other four conditions (with an average CRS of 5.755, which is close to neutral).

Table 4 The ANOVA results for CDT and CRS under non-yielding condition

| Metrics | External appearance | Descriptives | | | ANOVA | | | |
|---|---|---|---|---|---|---|---|---|
| | | Sample size | Mean | S.E. | df | F | p | $\eta p^2$ |
| CDT | nyCV0 | 290 | 7.507 | 0.063 | (4, 1433) | 1.445 | 0.217 | 0.004 |
| | nyAV0 | 291 | 7.486 | 0.059 | | | | |
| | nyAV1 | 288 | 7.522 | 0.060 | | | | |
| | nyAV2 | 287 | 7.504 | 0.059 | | | | |
| | nyAV3 | 282 | 7.338 | 0.072 | | | | |
| CRS | nyCV0 | 290 | 5.828 | 0.142 | (4, 1433) | 24.882 | <0.001 | 0.065 |
| | nyAV0 | 291 | 5.742 | 0.138 | | | | |
| | nyAV1 | 288 | 5.684 | 0.142 | | | | |
| | nyAV2 | 287 | 5.767 | 0.139 | | | | |
| | nyAV3 | 282 | 7.351 | 0.152 | | | | |

**(2) Gaze-based metrics**

A two-way MANOVA was also conducted to examine the impact of external appearance and AOI location on the Number of Fixations and Average Duration of Fixations under non-yielding conditions. The results are shown in Table 5. For the metrics of fixation counts, the main effect of external appearance was not significant; however, the main effect of AOI location was significant ($F(2) = 3.342$, $p = 0.036$, $\eta p^2 = 0.002$). Pairwise comparisons based on the LSD test indicate that the number of fixations in the Grill and Hood area was higher than in the Windshield area, with a small effect size.

Table 5 The ANOVA results for gaze-based metrics under non-yielding condition

| Gaze Metric | Source | df | F | p | $\eta p^2$ |
|---|---|---|---|---|---|
| Number of fixations | External appearance | 4 | 1.137 | 0.337 | 0.002 |
| | AOI Location | 2 | 3.342 | 0.036 | 0.002 |
| | External appearance * AOI location | 8 | 1.765 | 0.079 | 0.005 |
| Average duration of fixations | External appearance | 4 | 3.852 | 0.004 | 0.005 |
| | AOI Location | 2 | 92.827 | <0.001 | 0.059 |
| | External appearance * AOI location | 8 | 2.851 | 0.004 | 0.008 |

For the metric of Average duration of fixations, both the main effects and interaction effects of external appearance and AOI location were significant (see Table 5). Fig. 13a and Fig. 13b respectively present the pairwise LSD comparisons for external appearance and AOI location. Concerning external appearance, participants exhibited longer fixation durations on nyCV0 compared to nyAV1 and nyAV3. Non-yielding AVs with an eHMI (nyAV3) received lower average fixation durations compared to nyCV0, nyAV0, and nyAV2. Regarding AOI location, participants showed significantly longer fixation durations on the Grill and Hood areas of the vehicles compared to the windshield area. Fig. 13c illustrates participants' fixation durations across various AOIs when facing non-yielding vehicles manipulated with different external appearances. In the Grill area, nyCV0 had the longest fixation duration, surpassing all AVs. In the Hood area, nyAV3 exhibited the shortest average fixation duration, significantly lower than other non-eHMI-equipped vehicles. Across Windshield areas, vehicles with different external appearances showed no significant differences.

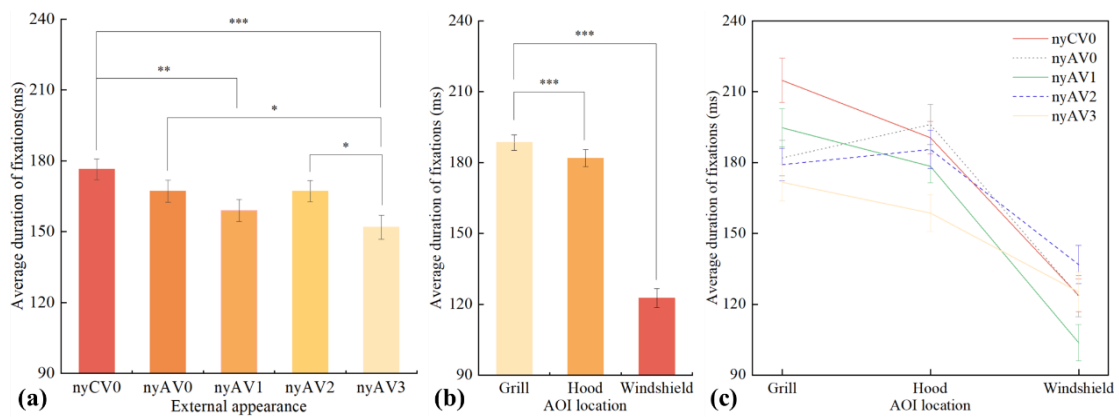

Fig. 13. Post-hoc comparisons and interaction effect for gaze-based metrics under non-yielding condition

## 4. Discussion

The current study aims to investigate how the exterior design of AVs affects pedestrians' subjective perceptions, visual behaviour, and road-crossing decisions under different movement patterns. Specifically, the study examines pedestrians' crossing behaviour when interacting with AVs, where the kinematic pattern (yielding, non-yielding) and external appearance (driver presence, textual indications, conspicuous sensors, and eHMIs) are manipulated. A video-based eye-tracking simulation experiment was conducted to collect multimodal pedestrian-vehicle interaction data, including participants' gaze metrics, road-crossing decision time, perceived clarity, and subjective evaluations of the AVs.

## 4.1 Pedestrians' trust in and attitude towards AVs

An intriguing finding is that participants' general trust in AV technology does not fully align with their attitudes towards crossing in front of an AV (see Fig. 7). When discussing AV technology, half of the participants exhibited a positive attitude, with nearly two-thirds of respondents believing that AVs can keep their lane and optimise the overall traffic system. However, when it came to crossing in front of an AV, participants became more cautious. Regarding the mental workload during crossing manoeuvres in front of an AV, the number of supporters and opponents was roughly equal (40%). This finding is consistent with the results of Rodríguez Palmeiro et al. (2018). Multiple factors may contribute to this outcome. For pedestrians, crossing in front of an AV presents a higher safety risk compared to walking alongside one. According to Thomas et al. (2018), crossing a roadway is the most common scenario for pedestrian collisions, exceeding the risk of walking along the roadway. Additionally, due to the limited availability of AVs on public roads, pedestrians may have no idea how AVs will interact with them. As indicated by Nordhoff et al. (2019), prior experience with AVs significantly enhances users' positive attitudes towards them. Similarly, after interacting with AVs equipped with eHMIs, pedestrians considered eHMIs important for AV-pedestrian interactions and crossing decisions (de Winter & Dodou, 2022).

## 4.2 The effect of vehicles' cues on pedestrian road-crossing decisions
### (1) Implicit cues

According to participants' responses to the post-experiment questionnaire (see Fig. 8), implicit cues of the vehicle played a predominant role in their crossing decisions. These cues included speed (M=4.48, SD=0.54), yielding behaviour (M=4.60, SD=0.64), distance to pedestrian (M=4.28, SD=0.76), and time-to-arrival (M=3.80, SD=0.89). The importance scores for these implicit cues were higher than for all explicit cues, with smaller variations. This finding aligns with previous studies on pedestrian-AV interactions (Lyu et al., 2024; Madigan et al., 2023).. A possible explanation is that these implicit cues directly affect pedestrian safety, whereas the impact of external features is indirect or less apparent (Rodríguez Palmeiro et al., 2018). Moreover, the multimodal data collected in this study provided unique, supplementary, and supportive evidence for this perspective. For instance, the distribution of pedestrians' road-crossing initiation distances varied significantly between the yielding and non-yielding conditions (see the yellow bars versus blue bars in Fig. 9). By contrast, within the yielding or non-yielding groups, the proportion of pedestrians' road-crossing percentage under the five different exterior manipulations was relatively consistent. This suggests that the vehicle's yielding behaviour predominantly determined whether pedestrians crossed in front of or behind the vehicle. Additionally, the descriptive analysis of pedestrians' road-crossing decision times (see Fig. 10) conveyed a similar narrative.

**(2) Textual indications**

Despite textual indications being rated highly by most participants (68%) in terms of subjective importance (M=4.05, SD=1.14), statistical analysis revealed that this exterior manipulation did not significantly affect participants' crossing decision times for both yielding and non-yielding AVs (see Fig. 11a and Table 4). This finding is in line with Rodríguez Palmeiro et al. (2018), who found that textual signs were self-reported by 70% of participants to have influenced their road-crossing decisions, yet had no significant influence on their actual gap acceptance. Furthermore, Fig. 11b shows that yielding AVs with textual indications (yAV1) received significantly lower perceived clarity compared to yCV0, yAV2, and yAV3. This suggests that pedestrians may become uncertain when encountering a vehicle labelled as an AV without providing additional information (e.g., Fig. 1b). Rodríguez Palmeiro et al. (2018) also noted that some participants in their post-experiment interviews expressed uncertainty about AVs with textual signs compared to CVs. Given that only 45% of respondents in this study considered it safe to cross in front of an AV, the high importance attributed to textual indications in crossing decisions may stem from pedestrians' scepticism or mistrust towards AVs.

**(3) Conspicuous sensors**

The questionnaire revealed that less than half of the respondents (47%) considered the roof radar system important in their crossing decisions (M=3.50, SD=1.29). Statistical analysis indicated that the external roof radar did not significantly affect pedestrians' crossing decision times for either yielding or non-yielding AVs (see Fig. 11a and Table 4). The external radar only slightly enhanced pedestrians' perceived clarity in the yielding condition (see Fig. 11b) compared to yAV1. Ackermans et al. (2020) found that rooftop sensors enhanced the perception of automated driving capability and elicited more positive behavioural responses. However, overall, the external roof radar had a relatively weak impact on pedestrians' subjective perceptions and crossing decisions. This could be attributed to the visual patterns pedestrians use when interacting with vehicles (road surface-grill-hood-windshield) (Bindschädel et al., 2022; Dey et al., 2020; Dey et al., 2019). In this experiment, in only 8.12% of all valid trials, participants' gaze behaviour was recorded in the roof AOI (covering and exceeding the radar's range). This suggests that pedestrians are less likely to focus on the roof of the vehicle when crossing, unless they are waiting for a taxi.

**(4) eHMI**

In this study, the advantages of eHMIs are multifaceted, as suggested by de Winter & Dodou (2022). Firstly, according to questionnaire responses, 65% of participants deemed eHMIs important in their crossing decisions (M=3.82, SD=1.18). Secondly, descriptive analysis of pedestrians' CID indicated that the implementation of eHMIs on AVs advanced pedestrians' CID, both in yielding and non-yielding conditions (see Fig. 9e and Fig. 9j). Additionally, for yielding AVs, eHMIs in the form of dynamic pedestrian icons significantly enhanced pedestrians' perceived clarity (M=8.73, close to the maximum score of 9) and reduced crossing decision times by over 1 second, compared to other exterior manipulations (see Fig. 11). The augmentation of subjective perception and reduction in decision times attributable to eHMIs in pedestrian-AV interactions have been consistently evidenced by extensive research (Bazilinskyy et al., 2019; Lee et al., 2022). Furthermore, eHMIs comprising pedestrian icons have been validated across multiple studies for their intuitiveness, comprehensibility, and clarity (Guo et al., 2022; Lyu et al., 2024). Regarding non-yielding AVs, eHMIs in the form of flashing AV icons did not significantly differ in CDT compared to other groups but did achieve significantly higher perceived clarity. This observation underscores the predominant role of implicit cues in pedestrian crossing behaviour.

**(5) Driver presence**

The importance of the presence of a human driver was rated lower (M=3.22, SD=1.24). Moreover, no significant differences were observed in pedestrians' crossing decision times and perceived clarity under AV0 and CV0 conditions. This finding is in line with Nuñez Velasco et al. (2021), indicating that the presence or absence of a human driver does not notably influence pedestrians' crossing decision time or safety margin. This is possible due to the fact that pedestrians tend to rely more on vehicle-based implicit information rather than driver-related explicit communication (Lee et al., 2021). However, some studies did suggest that informal communication from the driver, such as head movement, eye contact, or smiling, can enhance pedestrians' perceived clarity and safety (Onkhar et al., 2022). In the present study, the driver in CV0 did not exhibit such actions or expressions, representing a limitation of this research.

**4.3 The effect of AV's external appearance on pedestrians' gaze behaviour**

In this experimental study, two gaze metrics—Number of fixations and Average duration of fixations—were analysed to evaluate pedestrians' interactions with AVs featuring various exterior designs. Statistical analysis revealed that AVs' novel appearance significantly altered pedestrians' visual patterns towards approaching AVs. This effect was particularly spotlighted under yielding conditions (see Fig. 12).

Regarding the Number of fixations, significantly more fixations were recorded in the Grill AOI of yAV3 and the hood AOI of yAV1 and yAV2 (see Fig. 12c). Note that textual AV indications were placed in the hood areas of yAV1, yAV2, and yAV3, while a dynamic pedestrian icon-based eHMI was placed in the grill area of yAV3. According to the technical documentation of the Tobii eye-tracker, fixation count typically reflects the importance or level of interest in the area of interest (Tobii., 2021). Therefore, it can be inferred that, compared to CV0 or AV0, the novel exterior features of the AVs attracted more visual attention from participants in the corresponding areas. Furthermore, compared to yAV1 and yAV2, participants' fixations on yAV3 were highest in the grill AOI and lowest in the hood and windshield areas. This is partly because the grill area of yAV3 featured a dynamic and coloured eHMI, which led to increased visual visits. However, the significant reduction in fixations on the hood and windshield areas of yAV3 may be attributed to the eHMI used in this experiment, which has been proven to have high clarity and intuitiveness, providing pedestrians with sufficient evidence to initiate a safe road-crossing manoeuvre. Previous eye-tracking studies have indicated that clear eHMI designs can reduce the number and duration of pedestrians' fixations on AVs (Guo et al., 2022; Lyu et al., 2024).

In terms of the Average duration of fixations, a notable difference was observed that yCV0 received the longest average fixation duration in the Windshield AOI, significantly higher than AVs without drivers (see Fig. 12f). This may be due to the experiment including only a small number of trials (20%) with a human driver, serving as a conventional vehicle. Therefore, when a CV appeared with a driver in red, participants' visual attention were likely to be re-directed. Meanwhile, fixation duration on a specific AOI is generally associated with the complexity of the AOI and the mental effort of the viewer (Lyu et al., 2024; Meghanathan et al., 2015). In this case, as noted earlier, a limitation of this experiment was that the drivers maintained a fixed, static posture throughout, which might contradict pedestrians' real-world experiences, potentially leading to confusion or misunderstanding. Additionally, the average fixation duration on the hood areas of yAV1 and yAV2 was slightly higher than that of yAV3. This might be attributed to the textual indications on the AVs confusing the participants. This can be reflected in Fig. 11b, where participants' perceived clarity of yAV3 was significantly higher than that of yAV1 and yAV2. Eisma et al. (2023) found a similar effect in that flashing eHMI captured attention briefly, while the static text-based eHMI held attention for a longer period. Liu et al. (2023) indicated that pedestrians may tend to fix their gaze at a certain explicit cue longer when they do not accurately understand it.

Hence, an interesting conclusion could be made that in comparison to AV0 (devoid of a driver or additional external elements), AVs featuring textual indication (AV1) or a roof radar system (AV2) are likely to consume a greater share of pedestrians' visual resources. This is evidenced by heightened fixation counts and prolonged fixation durations (see Fig. 12), even leading to diminished perceptual clarity among pedestrians (see Fig. 8). However, the strategic deployment of a clear eHMI, such as in the form of pulsing pedestrian icons, serves to mitigate these adverse effects. Such an approach holds promise in reducing pedestrians' visual load, enhancing their perceptual clarity, expediting road-crossing decisions, and thereby, enhancing overall traffic efficiency. On the other hand, it also shows that excessive abuse of static text-based cues may run the risk of increasing pedestrian cognitive load.

Finally, under both yielding and non-yielding conditions, we observed a consistent pattern in visual fixation distribution across AOIs: participants exhibited significantly more fixations on the grill and hood areas than on the windshield. This finding aligns with the findings of Guo et al. (2022). Researchers have also argued that the more importance of an AOI, the more fixations it receives (Mahanama et al., 2022). This indicates that the grill and hood areas play a crucial role in pedestrians' perception of the vehicle's status and intent. Dey et al., (2019) observed that at a far distance, pedestrians typically judge a vehicle's behaviour and intent based on the motion of its front parts (such as the bumper, grill, and hood), while at closer range, they confirm their judgement by observing the driver through the windshield. This highlights the importance of implicit cues from the vehicle body. This also supports the grill area as an alternative to placing an eHMI for future AVs. An eHMI located on the bumper or grill may facilitate earlier recognition of the vehicle's intent by pedestrians, given the essential importance of this area.

### 4.4 Theoretical and safety implications

The current research has important implications in the following aspects:

(1) Our research is crucial for understanding the impact of the exterior design of AVs, correlated with its yielding patterns, on pedestrians' visual and road-crossing behaviours, particularly as they are about to undergo extensive public road testing. Previous studies suggested that AVs' novel textual signs and radar systems influenced pedestrians' subjective evaluations, rather than their gap acceptance or crossing initiation time. We discovered that the potential appearances of AVs altered pedestrians' visual attention patterns, such as where they looked and for how long. This change was more pronounced when the AV displayed a yielding intention.

(2) Our findings reveal that an eHMI can mitigate the visual load and perceptual confusion induced by AV's identity features (e.g., textual indications or radar sensors). Overall, pedestrians did not feel particularly safe when crossing in front of an AV. The AV's exterior identities, such as textual indications or radar systems, received lower perceptual clarity ratings and consumed more visual resources. However, the use of an eHMI alleviated these issues, reduced pedestrians' decision-making time, and enabled them to initiate crossing from an earlier distance. On the other hand, it shows that we need to pay attention to simplifying the design of static conspicuous explicit cues to reduce their occupation of pedestrians' visual and cognitive resources and avoid the risk of causing potential pedestrian decision delays.

(3) From a visual perception perspective, our findings demonstrate the potential of eHMI to enhance interaction efficiency and safety. It is commonly believed that dynamic, colourful, luminous components attract visual attention, such as traffic lights. Therefore, eHMI might be expected to occupy more visual resources, potentially diminishing attention to other traffic elements or road users. However, our study found that a well-designed eHMI did not increase the overall number of fixations or fixation duration for pedestrians and was able to convey the AV's yielding or non-yielding intention quickly and effectively. This provides visual perception evidence supporting the role of eHMI in ensuring pedestrian-vehicle interaction safety and efficiency. These findings are crucial for the future design of communication between AVs and pedestrians.

(4) This study, based on information processing theory (IPT), offers a new paradigm for research on pedestrian-AV interactions. According to IPT, pedestrians' behaviour when crossing in front of a vehicle involves information gathering, processing, and behavioural output, also known as "perception-cognition-action" (Predhumeau et al., 2021). However, existing studies primarily focus on the "action" stage of pedestrian-AV interaction, neglecting how pedestrians visually and auditorily perceive and cognitively process environmental and vehicular information. Our research is now advancing efforts to unveil this black box.

## 4.5 Limitations

This study has several limitations that need to be acknowledged. (1) The experimental study recruited a relatively young sample, considering that current road traffic accidents are the leading cause of death among young people (WHO, 2023). However, research indicates differences in visual patterns between older and younger age groups when crossing the street (Zito et al., 2015). Therefore, future studies should expand to include a broader age range. (2) The study used simplified traffic scenarios, i.e., one manipulated vehicle approaching on a two-way, two-lane road. Pedestrians' visual allocation may vary in more complex scenarios, for example, when interacting with more than one AV (Lyu et al., 2024). (3) The study only involved two kinematic patterns, yielding and non-yielding. Considering the dominant role of implicit cues in pedestrian road-crossing decisions, future research should explore pedestrians' visual and behavioural responses under more combinations of implicit and explicit conditions. (4) Due to the scarcity of AVs, the risk of pedestrian-AV interaction, and the convenience of manipulating AV appearance, this experiment adopted a video-based simulation paradigm. Although meaningful conclusions and implications were drawn, pedestrians' situational awareness or risk perception may be lowered by this experimental paradigm, thus affecting their information perception and decision-making processes. Therefore, the results of this study need to be validated in future real-world studies.

## 5. Conclusion

To investigate the impact of the exterior design of AVs, correlated with their kinematics, on pedestrians' subjective perception, visual attention, and road-crossing decisions, a video-based eye-tracking experiment was conducted. Multi-method data were collected on pedestrian gaze behaviour, perceived clarity, road-crossing initiation distance, and decision time during interactions with AVs, whose yielding patterns and external appearances were manipulated. The following conclusions are drawn from this research. (1) This research underscores the primary and dominant role of AV kinematic profiles in pedestrians' decision-making processes during interactions. This is evidenced by various factors, including the initiation distance and decision time for crossing, their responses in post-experiment subject evaluations, and their visual distribution on the approaching AV. (2) Novel features in current and forthcoming on-road testing AVs, such as textual identity indications and roof radar systems, do not significantly impact pedestrians' crossing decision time. However, they do increase the visual resources required by pedestrians, as demonstrated by heightened fixation counts, prolonged fixation durations, and reduced perceptual clarity, particularly in yielding conditions. (3) This study confirms that eHMI significantly enhances the efficiency of pedestrian-AV interactions. Beyond textual indications and radar systems, the addition of eHMI notably reduces pedestrians' visual load, improves perceptual clarity, shortens decision time, and facilitates earlier crossing decisions. These findings provide significant implications for the future external appearance and interaction design of AV, considering the visual and behavioural characteristics of vulnerable road users.

## CRediT authorship contribution statement

**W. L.:** Conceptualization, Data curation, Formal analysis, Investigation, Methodology, Software, Validation, Visualization, Writing – original draft, Writing – review & editing, Funding acquisition, Project administration. **Y. C.:** Conceptualization, Methodology, Funding acquisition, Project administration, Writing – review & editing. **Y. D.:** Conceptualization, Methodology, Funding acquisition, Project administration, Supervision, Writing – review & editing. **J. L.:** Investigation, Methodology, Visualization. **K. T.:** Conceptualization, Investigation, Methodology, Writing – original draft, Writing – review & editing, Project administration.

## Declaration of competing interest

The authors declare that they have no known competing financial interests or personal relationships that could have appeared to influence the work reported in this paper.

## Data availability

Data will be made available on request.

## Acknowledgements

This work was funded by the Natural Science Research Project of Anhui Educational Committee (Grant Number: 2023AH050934 & 2023AH030023), the Project of Humanities and Social Science Fund of Ministry of Education of China (Grant Number: 23YJC630032), and Scientific Research Starting Foundation of Anhui Polytechnic University (Grant Number: 2022YQQ091), and Natural Science Foundation of Anhui Province (Grant Number: 2208085MG183).